\documentclass{article}
\usepackage{spconf,amsmath,graphicx,booktabs,multirow,subfigure,hyperref,pifont}

\def\x{{\mathbf x}}

\title{speech and noise dual-stream spectrogram refine network with speech distortion loss for robust speech recognition}
\name{Haoyu Lu$^1$, Nan Li$^{1,\ast}$\thanks{*Corresponding author}, Tongtong Song$^1$, Longbiao Wang$^1$, Jianwu Dang$^1$, Xiaobao Wang$^{1,\ast}$, Shiliang Zhang}
\address{$^1$Tianjin Key Laboratory of Cognitive Computing and Application,\\College of Intelligence and Computing, Tianjin University, Tianjin, China}
\begin{document}
\ninept
\maketitle
\begin{abstract}
In recent years, 
  the joint training of speech enhancement front-end and automatic speech recognition (ASR) back-end has been widely used to improve the robustness of ASR systems. Traditional joint training
methods only use enhanced speech as input for the back-end. However, it is difficult for speech enhancement systems to directly separate speech from input due to the diverse types of noise with different intensities. 
Furthermore, speech distortion and residual noise are often observed in enhanced speech, and the distortion of speech and noise is different.
Most existing methods focus on fusing enhanced and noisy features to address this issue.
In this paper, we propose a dual-stream spectrogram refine network to simultaneously refine the speech and noise and decouple the noise from the noisy input.
%
Our proposed method can achieve better performance with a relative 8.6$\%$ CER reduction.

\end{abstract}
\section{Introduction}
\label{sec:intro}

Automatic speech recognition system has been widely applied on mobile devices for human-machine communication. 
Recently, ASR systems with end-to-end neural network architectures have developed rapidly \cite{dong2018speech,watanabe2017hybrid} and achieved promising performance.
Although significant progress has been achieved in ASR on clean speech, the performance of ASR systems is still far from desired in realistic scenarios.
There are various types of noise with different Signal-to-Noise Ratios(SNRs), which will sharply degrade the performance of the ASR systems. Thus, speech recognition in realistic scenarios remains
a considerable challenge

Robust speech recognition has been widely studied to improve the performance of ASR under complex scenes \cite{zhang2021end,chen2022noise,mendelev2021improved}. In \cite{prasad2021investigation}, they made an investigation of end-to-end models for robust speech recognition. 
There are two mainstream methods for robust speech recognition. 
One method is to augment the input data with various noises and reverberation to generate multi-condition data \cite{park2019specaugment,peso2022pmct}. Subsequently, the augmented data is fed to the end-to-end model.
Another method is to preprocess the input speech with speech enhancement techniques.
The existing works mainly use a two-stage approach to train a robust speech recognition model. The input speech is first passed through a speech enhancement (SE) module, and the enhanced speech is subsequently passed through an end-to-end speech recognition model.
 \begin{figure}[t]
  \begin{minipage}[b]{1.0\linewidth}
    \centering
    \centerline{\includegraphics[scale=0.8]{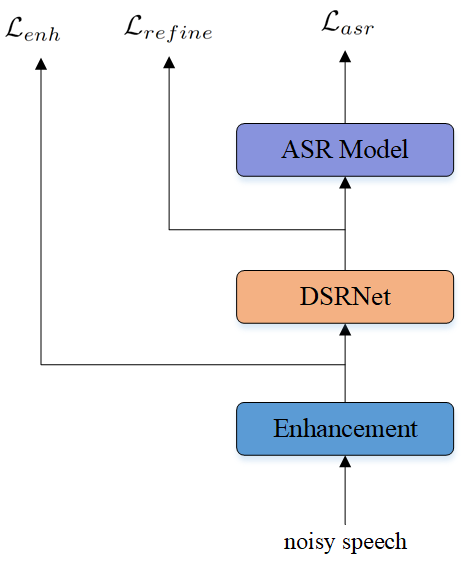}}
  \end{minipage}
    \caption{Block diagram of joint training framework.}
    \label{overview}
\end{figure}
Existing work \cite{weninger2015speech} has shown that Long Short-Term Memory (LSTM) RNNs can 
be used as a front-end for improving the noise robustness of robust ASR system. 

Joint training of the SE front-end and ASR back-end has been investigated to improve ASR performance \cite{ma2021multitask}. However, the speech enhancement based on deep neural network
often introduces speech distortion and remains residual noise which
may degrade the performance of ASR models. In \cite{iwamoto22_interspeech}, they investigate 
 the causes of ASR performance degradation
  by decomposing the SE errors into noise and artifacts. To alleviate the speech 
  distortion, \cite{hu2022interactive,zhuang2022coarse,fan2020gated}
  dynamically combined 
  the noisy and enhanced features during 
  training. In \cite{wang2020voicefilter}, they
 investigate the over-suppression problem. \cite{narayanan2022mask} presents a 
  technique to scale the mask to limit speech distortion
  using an ASR-based loss in an end-to-end fashion. In \cite{shi2021spectrograms},
  they propose a spectrogram fusion (SF)-based end-to-end robust ASR system, 
  in which the mapping-based and masking-based SE is simultaneously used as 
  the front end.
  In \cite{braun2022effect}, they provide insight into the advantage of magnitude
  regularization in the complex compressed spectral loss to trade off
  speech distortion and noise reduction. 

Although the existing joint training methods have greatly improved the robustness of ASR, there are still some problems. Specifically, the performance of ASR is affected by distortion or residual noise generated in SE. Few existing works have investigated effective methods for reducing distortion or residual noise. The main existing methods only fuse the distorted spectrogram with the original noisy speech features. There is still some residual noise in the fused features.
In this paper, we propose a speech and noise dual-stream spectrogram refine network (DSRNet) to estimate speech distortion and residual noise. We build the DSRNet to post-process the enhanced speech. Instead of only predicting the source speech and ignoring the noise features, we reuse the predicted features to refine the speech and noise simultaneously and decouple the noise features from the noisy input. 
We introduce a weighted MSE-based loss function that controls speech distortion and residual noise separately.
\begin{table}
  \small
  \centering
  \caption{The data structure of our dataset.}
  \scalebox{0.95}{
  \begin{tabular}{c|c|c}
    \toprule
     Subset & SNR & Noise corpus\\ 
     \midrule
     Training & \begin{tabular}{c} randomly selected from\\ {[-10, -5, 0, 5] dB} \end{tabular} & \begin{tabular}{c} 100 Nonspeech\\ Sounds \end{tabular}\\
     \midrule
     Development & \begin{tabular}{c} randomly selected from\\ {[-10, -5, 0, 5] dB} \end{tabular} & \begin{tabular}{c} 100 Nonspeech\\ Sounds \end{tabular}\\
     \midrule
     Test & \begin{tabular}{c} -10, -5, 0, 5 and\\ random dB \end{tabular}& \begin{tabular}{c} 100 Nonspeech\\ Sounds \end{tabular}\\
    \bottomrule
  \end{tabular}}
    \label{table1}
\end{table}
\vspace{-8pt}
\section{PROBLEM FORMULATION}
\label{sec:format}
Speech enhancement aims to remove the noise signals and estimate the target speech 
from noisy input. The noisy speech can be represented as:
\begin{equation} \label{eq1}
	{\rm y(t) = s(t) + n(t)} 
\end{equation}
where y(t), s(t), and n(t) denote the observed noisy signals, source signals and 
noise signals, respectively. We use the X (t, f), S(t, f) and N(t, f) as the 
corresponding magnitude spectrogram of noisy, source and noise signals, which 
still satisfy this relation:
\begin{equation} \label{eq2}
	{\rm Y(t,f) = S(t,f) + N(t,f)} 
\end{equation}
where (t,f) denotes the index of time-frequency(T-F) bins. We omit the (t,f) in 
the rest of this paper. The noisy magnitude spectrogram Y is used as the input of 
speech enhancement network. We formulate speech enhancement task as predicting
 a time frequency masks between noisy and clean spectrogram. 
 The conventional speech enhancement can be represented as follows:
\begin{equation} \label{eq3}
    \begin{aligned}
	& {\rm M = SE(Y)}  \\
	& {\rm \hat{S} = M \odot Y} \\
	\end{aligned}
\end{equation}
\vspace{-8pt}
\begin{equation} 
    \mathcal{L}_{enh} ={\rm MSE(\hat{S},S)}
  \end{equation}
where M is the estimated mask, ${\rm \hat{S}}$ is the estimated magnitude spectrogram of
source signals and $\odot$ denotes element-wise multiplication.
Speech distortion or residual noise are often observed in the enhanced speech.
We assume that there are still high correlations between the enhanced speech 
 and predicted noise features.
 We consider the predicted magnitude spectrogram of noise signal 
${\rm \hat{N}}$ is obtained by subtracting the enhanced signal  
${\rm \hat{S}}$ from the noisy signal Y. 
The predicted  spectrogram ${\rm \hat{S}}$ is composed of the source signal and the prediction error ${\rm E_{s}}$
 which is caused by speech distortion and residual noise. 
 And the predicted 
 spectrogram ${\rm \hat{N}}$ is composed of noise signal and the prediction error which may contain
 the missing information from source signal.
\begin{equation} 
  {\rm \hat{N}} = {\rm Y - \hat{S}}
\end{equation}
\vspace{-15pt}
\begin{equation} \label{eq8}
  {\rm \hat{S}} = {\rm S + E_{s}}
\end{equation}
\begin{equation} 
  {\rm \hat{N}} = {\rm N + E_{n}}
\end{equation}
\section{PROPOSED METHODS}
\subsection{Network Architecture} 
In this section, we discuss the details of the proposed method. 
We propose a speech and noise dual-stream spectrogram refine network (DSRNet) with a 
joint training framework to reduce speech distortion and 
residual noise as 
shown in Figure \ref{overview}. First, we feed noisy magnitude 
spectrogram features to the LSTM mask-based SE module. Then the 
estimated magnitude spectrogram ${\rm \hat{S}}$ and the predicted 
noise magnitude spectrogram ${\rm \hat{N}}$ are fed to the DSRNet 
to generate the refined magnitude spectrogram. 
We then extract 80-dim Fbank features from the refined spectrograms as input to the ASR model.
\begin{figure}[t]
  \begin{minipage}[b]{1.0\linewidth}
    \centering
    \centerline{\includegraphics[scale=0.8]{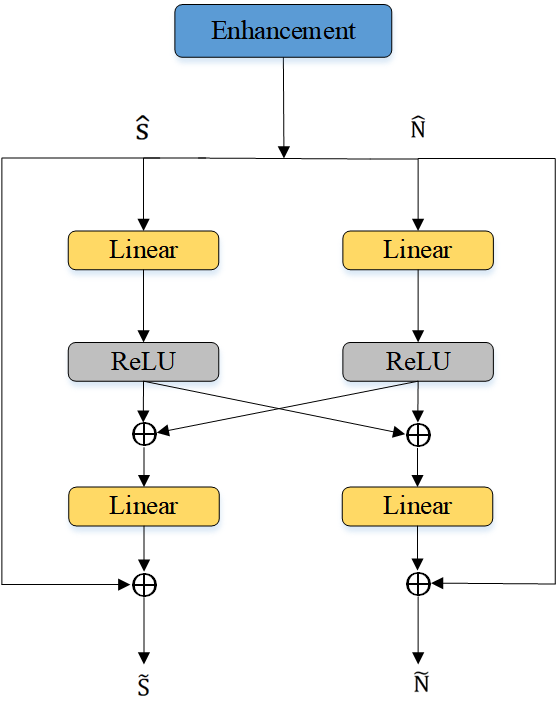}}
  \end{minipage}
  \caption{Block diagram of spectrogram refine network.}
    \label{NAWR}
\end{figure}
\vspace{-8pt}
\subsection{Spectrogram Refine}
\begin{table*}[htp]
  \centering
  \caption{CER results of the different methods on different test sets.}
  \scalebox{0.9}{
  \begin{tabular}{cccccccccccc}
  \toprule
  \multirow{2}{*}{Model} & \multirow{2}{*}{Joint Training} & \multicolumn{6}{c}{CER($\%$)}   & \multicolumn{1}{c}{\multirow{2}{*}{$ \alpha$}} & \multicolumn{1}{c}{\multirow{2}{*}{$\beta$}} & \multicolumn{1}{c}{\multirow{2}{*}{$\lambda$}}  & \multicolumn{1}{c}{\multirow{2}{*}{Paras(M)}} \\ \cline{3-8} 
                         & &-10dB & -5dB & 0dB & 5dB & Avg & \multicolumn{1}{l}{Random} & \multicolumn{1}{c}{} \\ \hline
  Transformer            &  &39.7     & 26.7     & 19.3    &  14.8   & 25.13    & 23.9 & —&— &— &16.67 \\
  SE+Trans            & No &46.2     & 30.5     & 21.3    &  15.4   & 28.35    & 26.3 &—& —&— &30.59 \\
  SE+Trans               & Yes &35.0     & 23.1    & 16.6    &  13.0   & 21.93    & 20.8 &300 & — & — &30.59 \\
  SE+DSRN+Trans               & Yes & 32.6    & 21.4    & 16.0     &  12.7    & 20.68     & 19.7    &300 & 0 & — &30.85 \\  
  SE+DSRN+Trans               & Yes &31.8     & 21.2    & 15.6    &  12.6   & 20.30    &   19.6 &300 & 100 & 0.5 &30.85 \\
  SE+DSRN+Trans        & Yes &\textbf{31.4}     & \textbf{20.7}     & \textbf{15.4}    & \textbf{12.6}    & \textbf{20.03}    & \textbf{19.2} &300 & 100 & Eq.\ref{11} &30.85 \\ \bottomrule
  \end{tabular}}
    \label{table2}
 \end{table*}
Figure \ref{NAWR} shows the block diagram of the proposed speech and noise 
dual-stream spectrogram refine network.
One stream computes the residual value for speech, and the other stream computes the residual value for noise.
Then we use the residual values to refine the enhanced speech and predicted noise, respectively.
\vspace{-10pt}
\subsubsection{Dual-stream Spectrogram Refine Network}
The dual-stream spectrogram refine network have two streams, 
one stream refines the enhanced speech and the other stream refines 
the predicted noise.
They share the same network structure but have separate network parameters.
We use DSRNet to compute the residual values, denoted as $\Theta$, which may contain over-suppression and missing information.
The residual values are added back to the spectrograms to counteract the undesired noise or speech and recover the distorted part.
The structure of the DSRNet is shown in Figure \ref{NAWR}.
 We can obtain the formulations as follows:
\begin{equation} 
  \begin{aligned}
  & {\rm \Theta_{s} = W_{\hat{s}}(W_{s} \hat{S} + W_{n} \hat{N}) + b_{\hat{s}}} \\
  & {\rm \Theta_{n} = W_{\hat{n}}(W_{s} \hat{S} + W_{n} \hat{N}) + b_{\hat{n}}}
  \end{aligned}
\end{equation}
\begin{equation} 
  \begin{aligned}
  & {\rm \tilde{S} = \hat{S} + \Theta_{s} }  \\
  & {\rm \tilde{N} = \hat{N} + \Theta_{n} }
  \end{aligned}
\end{equation}
where $\rm \Theta_{s}$ and $\rm \Theta_{n}$ are the residual values. 
$\rm W_{\hat{s}}$, $\rm W_{\hat{n}}$, $\rm W_{s}$, and $\rm W_{n}$ 
are the linear transformations, $\rm b_{\hat{s}}$ and $\rm b_{\hat{n}}$ are the bias terms.
\subsubsection{Weighted Speech Distortion Loss}

The distortion of the spectrogram is different for speech and noise.
Therefore, we propose a novel weighted speech distortion loss 
function in which both speech estimation error and 
noise prediction error are considered to overcome the distortion problem.
The loss function includes speech error term and noise error term. 
When the speech error is greater, 
we focus more on the speech. When the noise error is greater, we focus more on the noise. 
  \begin{equation}
  \begin{aligned}
    & {\rm E_{\tilde{s}} = \sum\nolimits_{t,f}|S - \tilde{S}| }\\
    & {\rm E_{\tilde{n}} = \sum\nolimits_{t,f}|N - \tilde{N}| }\\ 
  \end{aligned}
  \end{equation}
Therefore, the proposed mean-squared-error based loss function 
enables control of speech distortion 
and residual noise simultaneously. And the weighting $\lambda$ of 
loss function is time-varying between batches. Therefore, the loss function can be formulated as:
\begin{equation} \label{11}
\lambda ={\rm \frac{E_{\tilde{s}}}{E_{\tilde{s}}+E_{\tilde{n}}}}
\end{equation}
  \begin{equation} \label{eq6}
    \mathcal{L}_{refine} ={\rm \lambda MSE(\tilde{S},S) + (1-\lambda)MSE(\tilde{N},N) }
  \end{equation}
\label{sec:pagestyle}
\vspace{-10pt}
\subsubsection{Joint training}
We use a multi-task learning approach to jointly optimize the front-end and back-end to improve speech recognition performance. The loss function includes three terms. The weightings of speech enhancement and refine network loss are $\alpha$ and $\beta$, respectively.
\begin{equation} \label{loss}
  \mathcal{L} = \mathcal{L}_{asr} + \alpha \mathcal{L}_{enh} + \beta \mathcal{L}_{refine}
\end{equation}
\section{EXPERIMENTS}
\subsection{Dataset}
Our experiments are conducted on the open-source Mandarin 
speech corpus AISHELL-1\cite{bu2017aishell}. AISHELL-1 contains 400 speakers 
and more than 170 hours of Mandarin speech data. The training set contains 120,098 
utterances from 340 speakers. The development set contains 14,326 
utterances from 40 speakers. The test set contains 7176 utterances 
from 20 speakers. We manually simulate noisy speech on the AISHELL-1 with 100 Nonspeech 
Sounds noise dataset. We use the noise dataset to mix with the clean data of 
AISHELL-1 with 4 different SNRs each -10dB, -5dB, 0dB and 5dB. And we generate 
four SNRs test sets, and a random SNR test set. The simulated data and noise data
are released on Github\footnote
{https://github.com/manmushanhe/DSRNet-data}
for reference. The details are shown in Table \ref{table1}.
\vspace{-7pt}
\subsection{Experimental Setup}
In the experiments, we implemented a joint training system using the 
recipe in ESPnet\cite{watanabe2018espnet} for AISHELL. 
The input waveform is converted into STFT domain using a 512 window 
length with 128 hop length. The learning rate is set to 0.001 and 
the warm-up step size is set to 30, 000. In the SE module, 
the number of layers in the LSTM is two and the hidden size is 1024. 
In the DSRNet, the input and output sizes of the linear 
layers are 257. We use a Transformer based ASR model with 
80-dim Log-Mel features as input to the back-end. Hyperparameters 
$\alpha$ and $\beta$ are 300 and 100 respectively. 
For fair comparison, the training epoch is set to 70 for all experiments.
When the strategy of joint training is not used, we first pre-train the speech 
enhancement model, and then freeze the parameters of the
SE model to train the ASR back-end with the ASR loss.
\vspace{-7pt}
\subsection{Results}
\subsubsection{Evaluate the effectiveness of the proposed method}
We first compare our method with different models, and the results are 
shown in Table \ref{table2}. In Table \ref{table2}, 
``DSRN'' denotes our proposed speech and noise dual-stream spectrogram refine network.
``SE+Trans'' denotes the SE front-end and ASR back-end model.
As we can see, the performance of the jointly trained model can be significantly improved.
When there is no joint training, we first train a SE model using the magnitude spectral loss, and then freeze its parameters to train the ASR system.
The final objective of SE training and ASR training are different. SE is trained on the magnitude spectral loss and ASR is trained on the classification loss. 
There is a mismatch between SE and ASR. In the absence of joint training, the SE parameters cannot be tuned according to the ASR loss. 
The performance is bad. In the joint training, the SE network is not only learned to produce more intelligible speech,
 it is also aimed to generate features that is beneficial to recognition.

We conduct experiments to see how the result is affected by the weighting of 
SE loss in the joint training. Table \ref{table3} reports 
the average CER results on four SNRs test sets with different weightings of SE loss.
And we find that the weighting of SE loss has a significant impact on the result.
This may be because the loss of ASR is considerably greater than that of SE.
Experiments demonstrate that the performance of joint training model strongly depends 
on the relative weighting between each task's loss. 
When we continue to increase the value of $\alpha$, we find that the performance is no longer improving. 
There is an upper bound on the performance of 
using only SE networks with magnitude spectral loss. Therefore, we use the DSRNet with speech distortion loss
 to counteract the distortions and artifacts that are generated during SE.

Meanwhile, to evaluate the contribution 
of the DSRNet, we use the SE+Trans joint training model as baseline.
We set $\beta$ to 0 to investigate the impact of the speech distortion loss function. 
The results show that both the DSRNet and the loss function contribute to the performance.
And we set $\rm \lambda$ to 0.5 to investigate the impact of the weight $\lambda$ in Eq.\ref{11}.
The results show that the method of $\lambda$ computed based on Eq.\ref{11} is also effective.
Compared to the baseline approach, we achieve better performance with 
an average relative 8.6$\%$ CER reduction on four SNRs test sets, only at the 
cost of 0.26M parameters. 
The increase in model parameters and 
computations is slight. We experimented with other values of $\beta$, and in general, 
$\beta$ equal to 100 worked best, so we didn't show it in the table.
\begin{table}[t]
  \centering
  \caption{CER results with different weightings.}
  \scalebox{0.9}{
  \begin{tabular}{ccc}
  \toprule
  Model&{loss weightings($\alpha$)}&CER\_Avg(\%)\\ \hline
  \multirow{6}{*}{SE+Trans} &1&41.53\\
                              &50&22.10\\
                              &100&22.15\\
                              &200&21.98\\
                              &300&21.93\\
                              &400&21.90\\\bottomrule
  \end{tabular}}
  \vspace{-10pt}
    \label{table3}
 \end{table}
\vspace{-10pt}
\subsubsection{Visualization of Spectrograms}
\begin{figure}[t]
  \centering
  \subfigure[noisy]{
		\includegraphics[scale=0.2]{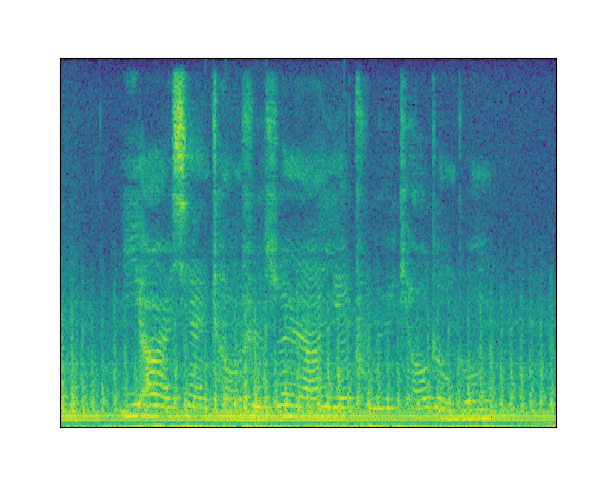}
	}
	\subfigure[clean]{
		\centering
		\includegraphics[scale=0.2]{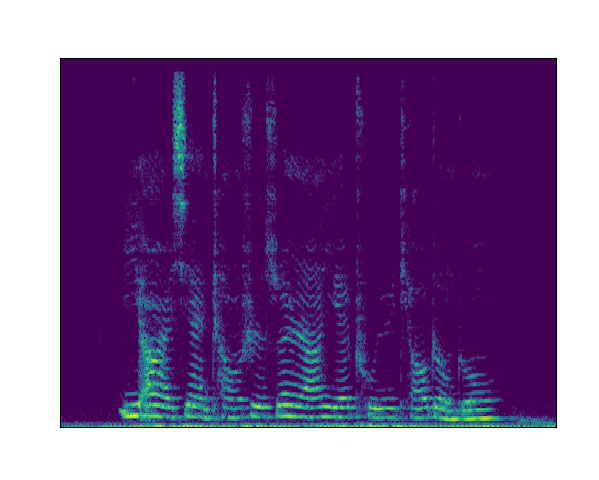}
	}
	\subfigure[baseline]{
		\centering
		\includegraphics[scale=0.2]{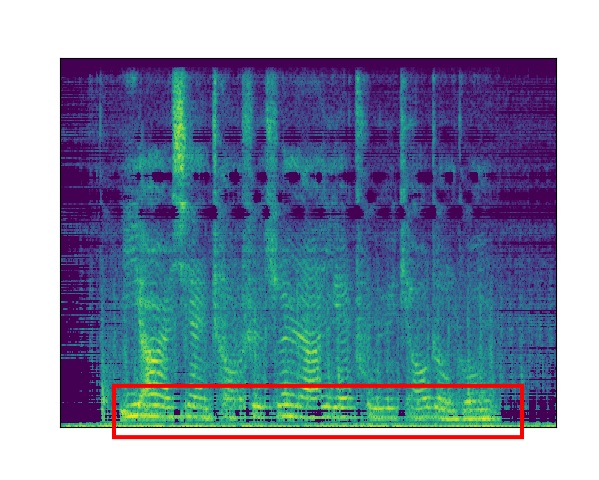}
	}
  \subfigure[enhanced-our]{
		\centering
		\includegraphics[scale=0.2]{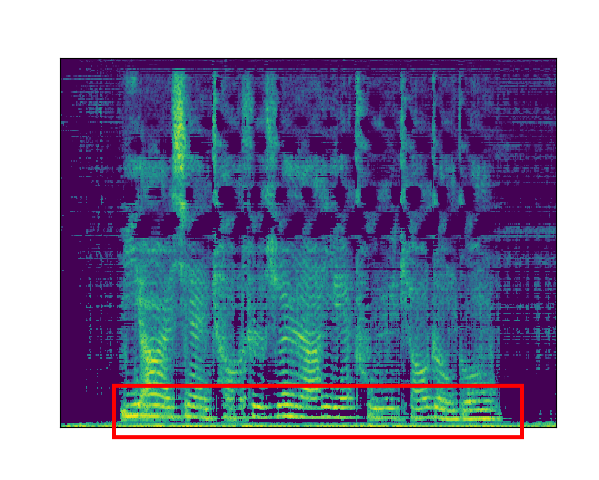}
	}
  \subfigure[refined-our]{
		\centering
		\includegraphics[scale=0.2]{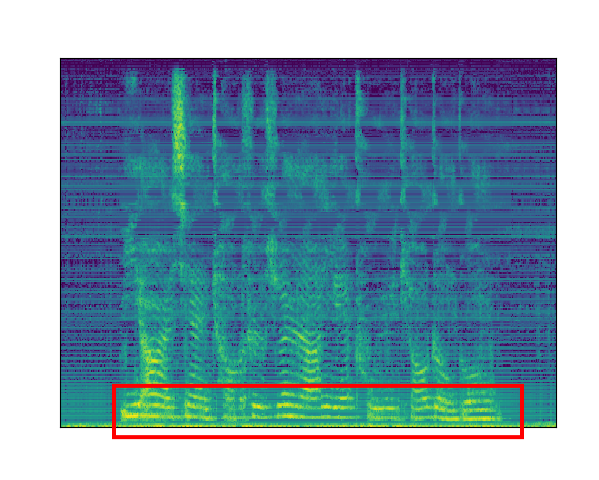}
	}
	\caption{Spectrograms from different methods.
	  \label{figure3}
  (a)noisy spectrogram.
  (b)clean spectrogram.
  (c)enhanced spectrogram from baseline.
  (d)enhanced spectrogram from our method.
  (e)refined spectrogram from our method.}
  \vspace{-10pt}
\end{figure}

In order to further understand how the DSRNet works, we visualize the
spectrograms from different model, as shown in Figure 
\ref{figure3}. (a) is the noisy spectrogram of simulated speech.
 (b) is the clean spectrogram of clean speech. (c), (d) is the enhanced 
 spectrogram in the baseline system and the enhanced spectrogram 
 in our method, respectively. And (e) is the refined spectrogram in our method. 

 Existing studies show that speech content is mainly
 concentrated in the low-frequency band of spectrograms.
 From (c) and (d), we see that part information in the 
 low-frequency band of enhanced spectrograms is missing and distorted, which means an over-suppression problem caused by SE.
Comparing (d) and (e), we can observe that the low-frequency band of the spectrogram
is refined. The enhanced spectrograms could recover some information in the 
 low-frequency band with the help of the DSRNet. This may mean the low-frequency band of spectrograms is more important for the ASR.
 These results show that the DSRNet indeed helps improve ASR performance and demonstrate the effectiveness of our method.
\vspace{-15pt}
\subsubsection{Reference Systems}
To evaluate the performance of the proposed method, we conduct experiments
 on four different systems for comparison. Table \ref{table4} shows 
 the results of reference systems.

 \textbf{Cascaded SE and ASR System}\cite{subramanian2019speech}: this paper jointly 
 optimizes SE and ASR only with ASR loss. 
 They investigate how a system optimized based on the ASR loss improves the speech enhancement 
 quality on various signal-level metrics. However, the results show that the cascaded system 
 tend to degrade the ASR performance. 

\textbf{GRF ASR System}\cite{fan2020gated}: they propose a gated recurrent fusion (GRF)
method with a joint training framework for robust ASR. The GRF unit 
is used to combine the noisy and enhanced features dynamically. We find that it 
performs worse in our experiments.

\textbf{Specaug ASR System}\cite{park2019specaugment}: they present a 
data augmentation method on the spectrogram for robust speech recognition.
The frequency mask and time mask are applied to the input of the ASR 
model, which help the network improve its modeling ability.

\textbf{Conv-TasNet and ASR System}\cite{von2020end}: this paper propose to combine 
a separation front-end based on Convolutional Time domain Audio Separation Network 
(Conv-TasNet) with an end-to-end ASR model. They jointly optimize the network with 
the Scale-Invariant-Signal-to-Noise Ratio (SI-SNR) loss and a multi-target loss for the ASR system.


\begin{table}[t]
  \centering
  \caption{CER results of different methods.}
  \scalebox{0.9}{
  \begin{tabular}{lc}
  \toprule
  Model                      & CER\_Avg(\%) \\ \hline
  Cascaded SE and ASR System & 29.05             \\
  GRF ASR System             &29.01              \\ 
  Specaug ASR System         &23.50              \\
  Convtasnet and ASR System  &22.35                     \\  \hline
  Our proposed System        &\textbf{20.03}             \\\bottomrule
  
  \end{tabular}}
  \vspace{-10pt}
    \label{table4}    
\end{table}


\vspace{-10pt}
\section{CONCLUSION}
In this paper, we explored the effect of weights of loss function on ASR performance. Experiment results show that the performance of joint training systems highly depends on the relative weights of each loss and the speech enhancement network will introduce speech distortion. We proposed a lightweight speech and noise 
dual-stream spectrogram refine network with a joint training framework for reducing speech distortion. The DSRNet estimate the residual values by 
reusing the enhanced speech and predicted noise, which can counteract 
the undesired noise and recover the distorted speech. We designed a weighted speech distortion loss to control of speech distortion and residual noise simultaneously.
Moreover, the proposed method is simple to implement 
and introduces a few computational overheads. 
Final results show that the proposed method performs better
 with a relative 8.6$\%$ CER reduction.
\vspace{-10pt}
\section{ACKNOWLEDGEMENTS}
This work was supported in part by the National Natural Science Foundation of
China under Grant 62176182 and Alibaba Group through the Alibaba
Innovative Research Program.
\clearpage
\vfill
\pagebreak
\bibliographystyle{IEEEbib}
\bibliography{strings,refs}

\end{document}